\documentclass{article}

\usepackage[nonatbib, final]{neurips_2019}

\usepackage[utf8]{inputenc} 
\usepackage[T1]{fontenc}    
\usepackage{hyperref}       
\usepackage{url}            
\usepackage{booktabs}       
\usepackage{amsfonts}       
\usepackage{nicefrac}       
\usepackage{microtype}      

\usepackage{graphicx}
\usepackage{float}
\usepackage{amsmath}
\usepackage{subcaption}
\usepackage{caption}
\usepackage{comment}
\usepackage{ulem}
\usepackage{xspace}
\newcommand{\TarLig}{\texttt{TarLig}\xspace}

\title{Leveraging binding-site structure for drug discovery with point-cloud methods}

\author{
  Vincent Mallet\\
  Department of Computer Science\\
  McGill University\\
  \texttt{vincent.mallet@mail.mcgill.ca} \\
  \And
  Carlos G. Oliver\\
  Department of Computer Science\\
  McGill University\\
  \texttt{carlos.gonzalezoliver@mail.mcgill.ca} \\
  \And
  Nicolas Moitessier\\
  Department of Chemistry\\
  McGill University\\
  \texttt{nicolas.moitessier@mcgill.ca} \\
  \And
  Jérôme Waldispühl\\
  Department of Computer Science\\
  McGill University\\
  \texttt{jerome.waldispuhl@mcgill.ca} \\
}

\begin{document}
\maketitle

\begin{abstract}



Computational drug discovery strategies can be broadly placed in two categories: ligand-based methods which identify novel molecules by similarity with known ligands, and structure-based methods which predict molecules with high-affinity to a given 3D structure (e.g. a protein). 
However, ligand-based methods do not leverage information about the binding site, and structure-based approaches rely on the knowledge of a finite set of ligands binding the target.
In this work, we introduce \TarLig, a novel approach that aims to bridge the gap between ligand and structure-based approaches. 
We use the 3D structure of the binding site as input to a model which predicts the ligand preferences of the binding site. 
The resulting predictions could then offer promising seeds and constraints in the chemical space search, based on the binding site structure. 
\TarLig outperforms standard models by introducing a data-alignment and augmentation technique. 
The recent popularity of Volumetric 3DCNN pipelines in structural bioinformatics suggests that this extra step could help a wide range of methods to improve their results with minimal modifications.

\end{abstract}

\section{Introduction}

\paragraph{\textit{In silico} drug discovery}
The majority of drugs are small organic molecules that alter cellular mechanisms upon binding to key bio-molecules such as proteins. 
Identifying molecules with a desired set of properties is known as the drug discovery process.  
This process often relies on an iterative, tedious and expensive process. 
Because most molecules are inactive or lack specificity toward a given function and binding site, Virtual Screening (VS) tools have emerged as an essential tool to identify the most promising candidates molecules prior to synthesis and/or experimental evaluation. 
Given a initial set of generic small molecules (i.e. the library), VS methods output a subset of this library enriched with active compounds. 
Another use of these computational methods is to predict off-target binding (binding to a non-desired site) to assess the binding specificity to the molecular target. 
This information is in particular helpful to anticipate potentially toxic compounds.

\paragraph{Structure-based approaches}
Most VS methods broadly fall under two categories \cite{ligands_vs_BP}. Structure-based approaches use the 3D structure of the protein binding site to identify promising candidate ligands (e.g., enzyme inhibitors, receptor antagonists,...). 
If the binding site is not known, several tools are able to detect the binding sites of a protein \cite{FPocket,Jimnez2017}. 
Knowledge of the binding site shape and requirements from ligand-protein co-crystals can be exploited to dock ligands into this binding site. 
This process is computationally demanding (i.e., from a few seconds to a few minutes per ligand) and consists in two phases: one that explores possible binding orientations of the ligands in the binding site and another that ranks them by affinity. 
This approach has been the subject of detailed studies \cite{Lang2009,Ballester2010}. 
Some machine learning approaches can be used to do the ranking step or even bypass the docking step and directly compute an affinity score\cite{Pereira2016}. 
However, increased accuracy comes at the cost of trying a larger number of orientations and therefore cannot be easily accelerated. 
The shortcoming of such methods is that they take as input both a ligand and the target protein and thus they are restrained to a limited set of ligands. 
This leads to limitations because this finite set has already been extensively studied and patented. 
Additionally, these methods are often computationally expensive.

\paragraph{Ligand-based approaches}
Ligand-based approaches assume that ligands structurally similar to active ligands will also be active\cite{Lyne2002,Hamza2012}. 
While this concept ignores the actual biological binding site explicitly, it implicitly considers the potential interactions and takes advantage of experimental binding affinities of reference ligands to search for new ones. 
Interest towards generative models in this ligand-centered task is rising\cite{Kadurin2017}. 
A major asset of this class of methods is the ability to identify potential ligands that lie outside of the patented molecular space. 
These models often rely on unsupervised methods that use a similar strategies to \texttt{word2vec} in the molecular space \cite{GomezBombarelli2018}. 
The latent space induced by such methods is conveniently prone to algebra. 
Notably, we can use optimization processes in the latent space to fine-tune a seed for ligand druggability \cite{Olivecrona2017, Elton2019DeepLF}. These methods are becoming increasingly popular but struggle to use binding site information, relying instead on extensive experimental assays to reference ligands. 
Although it is not a fundamental bottleneck for target prediction tasks, this is a major limitation for off-target predictions.

\paragraph{Our contribution}
In this paper, we propose to integrate binding site information into a ligand-based approach. 
For this purpose, we introduce a new task: we use protein structure information to quickly identify regions of interest in ligand space and use this knowledge as a potential guide in the ligand generative process. 
More specifically, we use the 3D structure of a protein binding site as input and predict a latent space embedding of a ligand. 
To our knowledge, this kind of approach was only conducted independently in a preprint \cite{DBLP:journals/corr/abs-1809-02032} that used graphs to represent the pocket.
Our model learns a rotation invariant representation for the 3D input using an alignment and augmentation step, showing superior results against 3D steerable and classic Volumetric CNNs pipelines. 
We believe this step is relevant for 3D shape comparison in general and could help other Volumetric CNN models improve their results.

\section{Related methods and data representation}
\subsection{Ligand representation}
\paragraph{Sparse representations}
Ligands can be represented as graphs, strings, sets of subgroups or 3D structures. 
All of these representations can in turn be embedded into vectors. 
 Molecular fingerprints are a commonly used representation defined as a bit-string encoding the presence or absence of certain relevant chemical fragments according to domain experts \cite{Durant2002, Rogers2010}. 
 Such representations have the usual advantages and drawbacks of hand-featured representations: they do not rely on abundant data and are easily interpretable. 
 However, they are also not compact and potentially miss important features of the chemical space. 
 The largest chemical compounds database (REAL database) contains $10^9$ compounds that could be represented using binary vectors of dimension $log_2(10^{9}) \sim 30$ instead of the fingerprints of length 1024 used by the ECPF6 \cite{Rogers2010} representations. Thus, hand-crafted, binary embeddings and the bit-to-bit distance (Tanimoto coefficient) span a sparse discontinuous space with regards to the space usually explored by chemists. 
 Continuous embeddings allow a more compact embedding and better preservation of chemical similarity that are the basis of all potential machine learning algorithms. 

\paragraph{Data-driven representations}
Data-driven representations arose in 2015 and were first applied to graph representations of ligands \cite{duvenaud2015convolutional, Kearnes2016}. Next, unsupervised methods of auto encoding were applied to larger data sets using the \texttt{SMILES} representation of millions of compounds available in large databases such as ZINC\cite{Irwin2012} with a paper leveraging similar ideas as \texttt{word2vec} to get embeddings of molecules \cite{GomezBombarelli2018}. The authors were the first to introduce the idea that the latent space could serve as a good molecular representation for other learning tasks such as bio-activity prediction.  A lot of work was conducted in this domain to adapt it to the use of VAEs or GANs\cite{Elton2019DeepLF} and we chose to base our work on a transitional auto encoder \cite{Winter2018} trained from \texttt{SMILES} to canonical \texttt{SMILES} on over one billion data points. 
We will represent our ligands with the latent space representation of this model.

\subsection{Protein representations}
The functional level of proteins and of most biological objects resides in their 3D structure and is composed of finite number of elements. 
We preferred the atomic resolution over the amino-acid resolution as the latter cannot properly model the orientation of each amino acid and side-chain details,  two major factors in protein-ligand binding. 
Therefore the binding sites are fundamentally modeled as a point cloud of atoms. 
Several deterministic embedding tools exist that turn these objects into vectors \cite{Batista2014} but for reasons similar to ligands, learned representations are more promising \cite{Kuzminykh2018}. 

\subsection{Point cloud networks}
While we chose to model protein binding sites as a point cloud - ie a matrix of coordinates associated with atom types - for biological motivations, there is no well established way to process such data. 
The challenges are choosing a translation and rotation invariant coordinate system to express our data in and an order to consider our points. 
The point ordering is partially solved by a framework such as PointNet \cite{Charles2017} using MaxPooling over all points. The most common solution to the ordering challenge is to consider this cloud as a sparse 3D image and to leverage the CNN frameworks, which consists in the Volumetric CNN framework. Treating objects as images fails to respect their rotational invariance and leads to very sparse inputs. For 2D images there is usually an orientation convention but there is no such native pose of a 3D object. 
This makes data augmentation much more difficult. Moreover as argued in \cite{Thomas2018TensorFN} the data augmentation is much more challenging in 3D than in 2D.

For these reasons, recent works try to extend the invariance properties of CNNs. 
In 2D, some work has shown superior performance including rotational and symmetry invariances \cite{Jaderberg:2015:STN:2969442.2969465,pmlr-v48-cohenc16}. 
In 3D, recent models manage to include rotational invariance\cite{Schtt2018} or equivariance\cite{Thomas2018TensorFN, DBLP:conf/nips/WeilerGWBC18}. 
These models have the ability to leverage the data properties. However they do not usually have clear convergence properties. Also since they have not been extensively used and studied, they are harder to use and to train than usual networks. We implemented a Volumetric pipeline with modifications and we chose to also use steerable 3DCNN\cite{DBLP:conf/nips/WeilerGWBC18} because they required only minimal modification from the Volumetric pipeline

\section{Methods}

\subsection{Data and performance metrics}
The protein binding sites we used were extracted from the PDB\cite{Berman2000} from bound examples. The extraction process consisted in taking bound protein structures, extracting at all amino-acids around the ligand and extending the radius around each of its atom until a certain radius or a certain number of neighbors is reached . We removed the metabolites in order to focus on larger, more interesting binding molecules. We then removed the binding sites that we considered duplicates using sequence identity cutoff and removing several examples of the same ligand bound to the same protein. 

\begin{table}[H]
  \caption{Data extracted from the PDB}
  \label{data-table}
  \centering
  \begin{tabular}{lll}
    \cmidrule(r){1-2}
Number of proteins & Size in Angstrom & Number of ligands \\
    \midrule
    30964         &    40 * 30 * 30    & 3362  \\
    \bottomrule
  \end{tabular}
\end{table}

As a metric we will often use the Mean Square Error (MSE) and the Enrichment factor (EF). EF at the $i^{th}$ level consist in the rate of active ligands in a subset of size $i\%$ of the original data set, normalized by the rate of actives in the whole set. EF was designed to mimic the true use case of virtual screening where chemists only test a small subset and hope to have as many actives as possible.

We used the DUDE \cite{Mysinger2012} database that is the standard for assessing enrichment factors.  The DUDE database consists in 102 targets and their associated set of actives. They then create 50 decoys per actives, preserving the physico-chemical features, but making the structure different, to compromise the binding.

\subsection{Principal axis alignment}
We also offer an alternate  strategy to bypass the problem of rotation invariance. For 3D objects embedded in images, we can compute the PCA as a way to find a consensus way to align these objects addressing the lack of native pose problem. Computing these axis is costly but it can be done as a preprocessing step and enables us to bypass the registration problem. Since the eigenvectors have no canonical sign, there remains an ambiguity regarding the pose but for a k-dimensional space, it is reduced to $2^k = 8$ possible poses for a 3D object.

We first validated this idea to use the principal axis of 3D objects by applying it to a small molecules comparison method : USRCAT \cite{Schreyer2012}. This method uses the successive first moments of the distribution of distances to references points as an embedding for a given molecule. As an example, the first feature of a set of points would be its mean distance to its centroid. We used the PCA-based approach to get orientation independent reference points (at a given distance in angstrom from the centroid, following the eigenvectors axis). We used the benchmark used in the latest version of this method and show superior performance in term of enrichment factor.

We then used this idea to represent our binding sites all aligned on the same grid. A side advantage of this alignment step is that it enables us to fit all object on a smaller grid since the dimensions with the highest variance usually correspond to the ones with largest values - with the exception of outliers. Using this framework solves the translation problem and enables us to drastically reduce the problem of representing rotations of point clouds. We can use data augmentation to put all 8 flipped views in the data set. We can then think our network might learn the invariance to flips. If we use a permutation invariant reduction operation such as the average on the flips, we make the network invariant to rotations. 

To explicitly enforce invariance to rotations for each prediction, we need to tie weights that are symmetric with respect to one of the axes of the 3D grid. 
We can also present batches containing the 8 poses and the batch-averaged gradients will then be symmetric. This is an engineering fix that makes the implementation straightforward. Further implementation of this model should include the weight tying. This would reduce both the batch size and the number of parameters. However, since the number of possible models is the same, we should be getting similar results with such constraints actively enforced.

This simple alignment could be an easy yet efficient engineering solution to help building rotation invariant networks. However this approach has a few caveats : we need to compute the PCA of the point cloud for every evaluation of the networks, which could be challenging for high number of points - but not for molecular applications. 

\begin{figure}[h]
  \centering
  \includegraphics[width=0.95\textwidth] {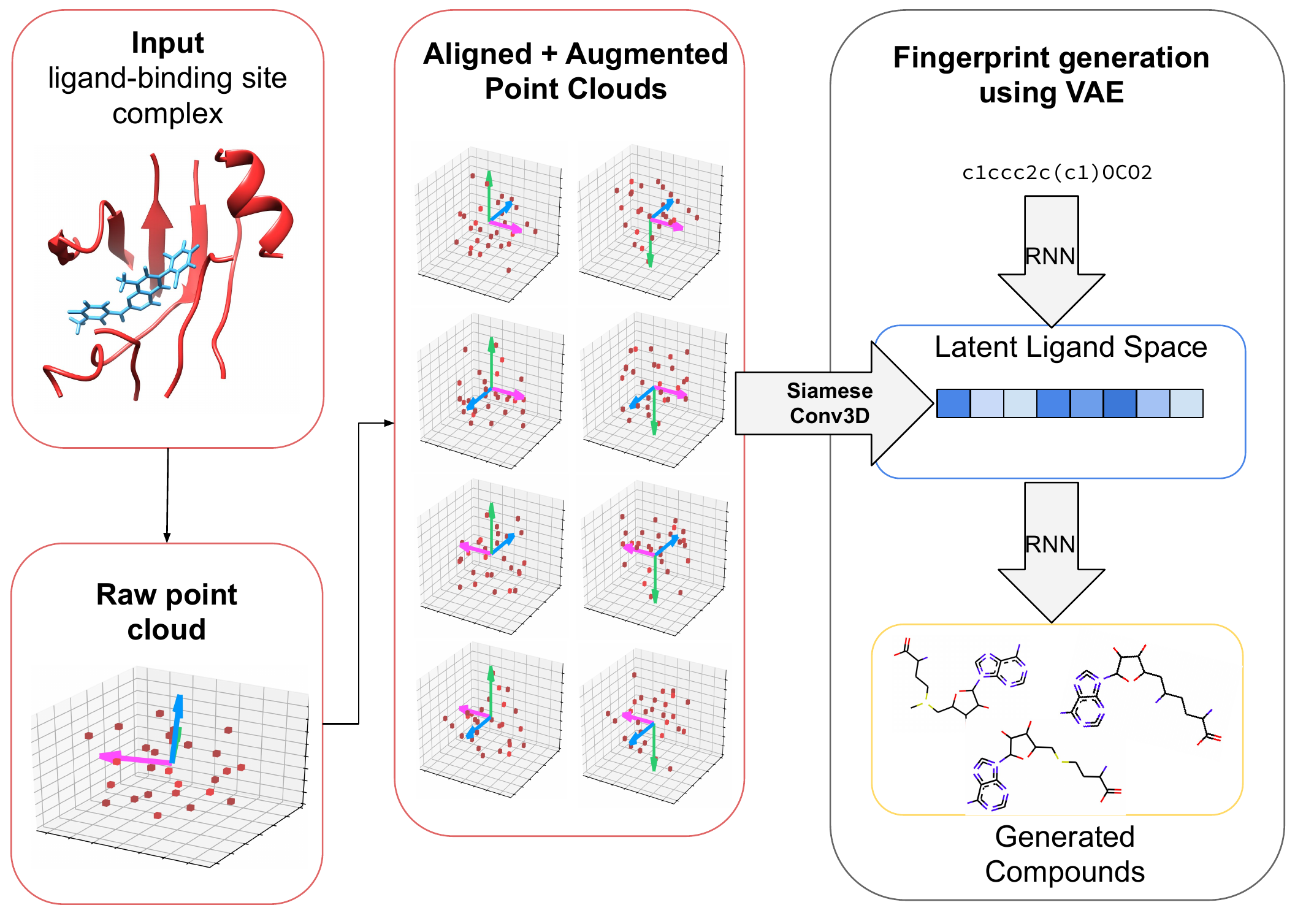}
  \caption{Pipeline used : a binding site (in red) is extracted around a ligand (blue) and turned into a point cloud. This point cloud is aligned to its principal axis and turned into 8 volumetric images. These representations are then used to learn the latent code of the extracted ligand }
  \label{pipeline}
\end{figure}

We used the pipeline presented in {\bf Figure \ref{pipeline}}.

\section{Results and discussion}

\subsection{Alignment Step Validation}

To assess the impact of the alignment step with a side application, we competed against USRCAT\cite{Schreyer2012} using reference points derived from the PCA. 
We computed EF at different thresholds for each of the DUDE\cite{Mysinger2012} binding sites and averaged the results. 
We obtain similar yet significantly superior results ($\sim 9$\%). 
The limited improvement can be explained by the method reaching its limits, as explained in \cite{Schreyer2012} but is enough to show that using points derived from the PCA is a promising solution for representing 3D objects.

\begin{table}[H]
  \caption{EF computed with different threshold averaged over the DUDE database (the ratio vs our method)}
  \label{EF}
  \centering
  \begin{tabular}{lll}
    \cmidrule(r){1-2}
    Method     & $\text{EF}_{0.01}$     & $\text{EF}_{0.0025}$ \\
    \midrule
    UFSR         &    7.97  (0.45)    &       4.50  (0.46)  \\
    USRCAT       &    16.50 (0.91)    &       9.14  (0.94)  \\
    USRPCAT      &    17.51    (1)    &       9.97 (1)      \\
    \bottomrule
  \end{tabular}
\end{table}

\subsection{3DCNN Results - Alignment step}
We now come back to the task of predicting latent representation of ligands from the structure of a binding site. 
We have trained a number of models to assess the impact of :
 \begin{itemize}
     \setlength\itemsep{0em}
     \item Aligning the data to the PCA axes
     \item Augmenting the data set with flips
     \item Using batching or siamese model to enforce invariance 
     \item Different CNN settings with more or less parameters
 \end{itemize}
See {\bf Section \ref{implementation}} for more details on the implementation and infrastructure architecture and parameters.

As a control experiment, we trained our model on shuffled labels. We benchmarked against steerable CNN implementations\cite{DBLP:conf/nips/WeilerGWBC18}. 
We also wanted to assess the bagging effect of models, so we show the average error on each view of a 3D image in addition to the bagged error.

\begin{table}[H]
  \caption{Mean square error (MSE) between true and predicted ligands.}
  \label{MSE}
  \centering
  \begin{tabular}{lllll}
    \cmidrule(r){1-2}
    Method  & Best test MSE & Bagged test MSE & Final train MSE & Time to train\\
    \midrule
    \TarLig shuffled       & 0.210  & 0.206 & 0.07  &  6h \\
    \TarLig                & 0.108  & 0.108 & 0.038 &  1h \\
    \TarLig flips          & 0.095  & 0.092 & 0.04  &  6h \\
    \TarLig PCA            & 0.103  & 0.103 & 0.036 & \textbf{1h} \\
    \TarLig PCA flips      & 0.089  & \textbf{0.085}& 0.033 & 5h \\
    \TarLig batched flips  & 0.095  & 0.088 & 0.055 & 5h \\
    \TarLig siamese        & \textbf{0.088} & 0.088 & \textbf{0.023} &  4h \\
    Small \TarLig flips       & 0.099  & 0.089 & 0.035 & 5h \\
    Se3cnn flips         & 0.18   & 0.16  & 0.091 & 35h\\
    Small se3cnn flips   & Diverged & N/A & 0.17  & 19h\\
    \bottomrule
  \end{tabular}
\end{table}

We see in {\bf Table {\ref{MSE}}} that our model is able to learn, and performs significantly better than the shuffled control.  
We have the expected result with regards to the PCA alignment : aligning all binding sites results in a higher score. 
Moreover, these results were obtained using early stopping, but considering the test metric curves, we see that the learning is a lot less stable with unaligned data, i.e. the model is harder to train and the best test error is actually over-fitting on the test set (data not shown). 
Therefore, we conclude that our alignment strategy is beneficial. 
The data augmentation also has the expected effect of helping the learning. 

The best model is the one that does not enforce the rotational invariance. 
Putting all flips in the same batch results in the same value of convergence but takes longer to reach it. 
This can be interpreted as an 'effective batch size' effect : we enforce rotational invariance at the cost of showing eight times less binding sites in each batch batch (for a constant batch size). 
The siamese model only pushes the bagged average to the right vector while the other push each individual prediction to the label.

All architectures we trained behaved very similarly. We report the \TarLig one that achieved the best, yet comparable, results to the ones we trained after, including smaller versions. The trends we just described concerning the impact of aligning the pocket, augmenting the data or using it in a siamese architecture were found to be true across all the models we trained.

The se3cnn network showed poor performance overall and we could not manage to bring its performance close to our results. 
We think it is a big limitation of such networks, despite having appealing theoretical properties, they are much harder to train and use which is a barrier to fully leveraging their power. 
The alignment step enabled learning for this task and we think that combined with the volumetric approach and some of the refinements developed for this approach, it could be an efficient and much easier way to use 3D data.

\subsection{Performance by ligand family}

\begin{figure}[H]
    \centering
    \begin{subfigure}[t]{0.45\textwidth}
        \centering
        \includegraphics[width=\textwidth]{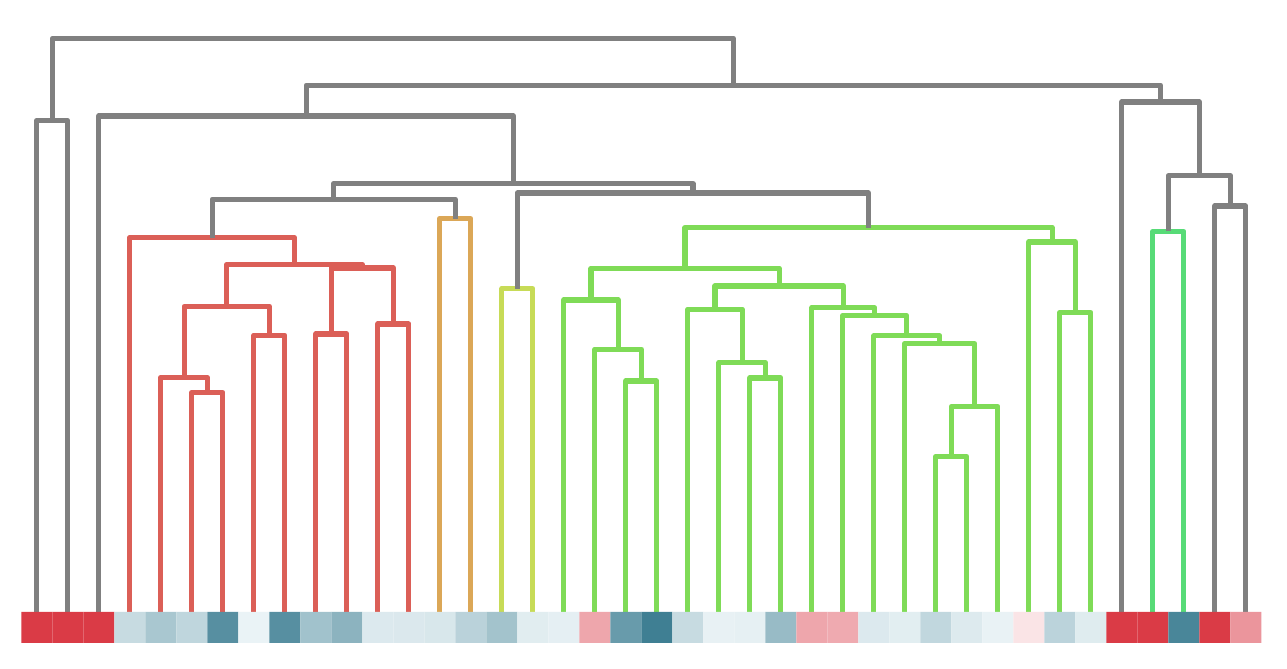}
        \caption{Performance by ligand, with clustering of the ligands}
    \end{subfigure}
    ~ 
    \begin{subfigure}[t]{0.45\textwidth}
         \centering
        \includegraphics[scale=0.35]{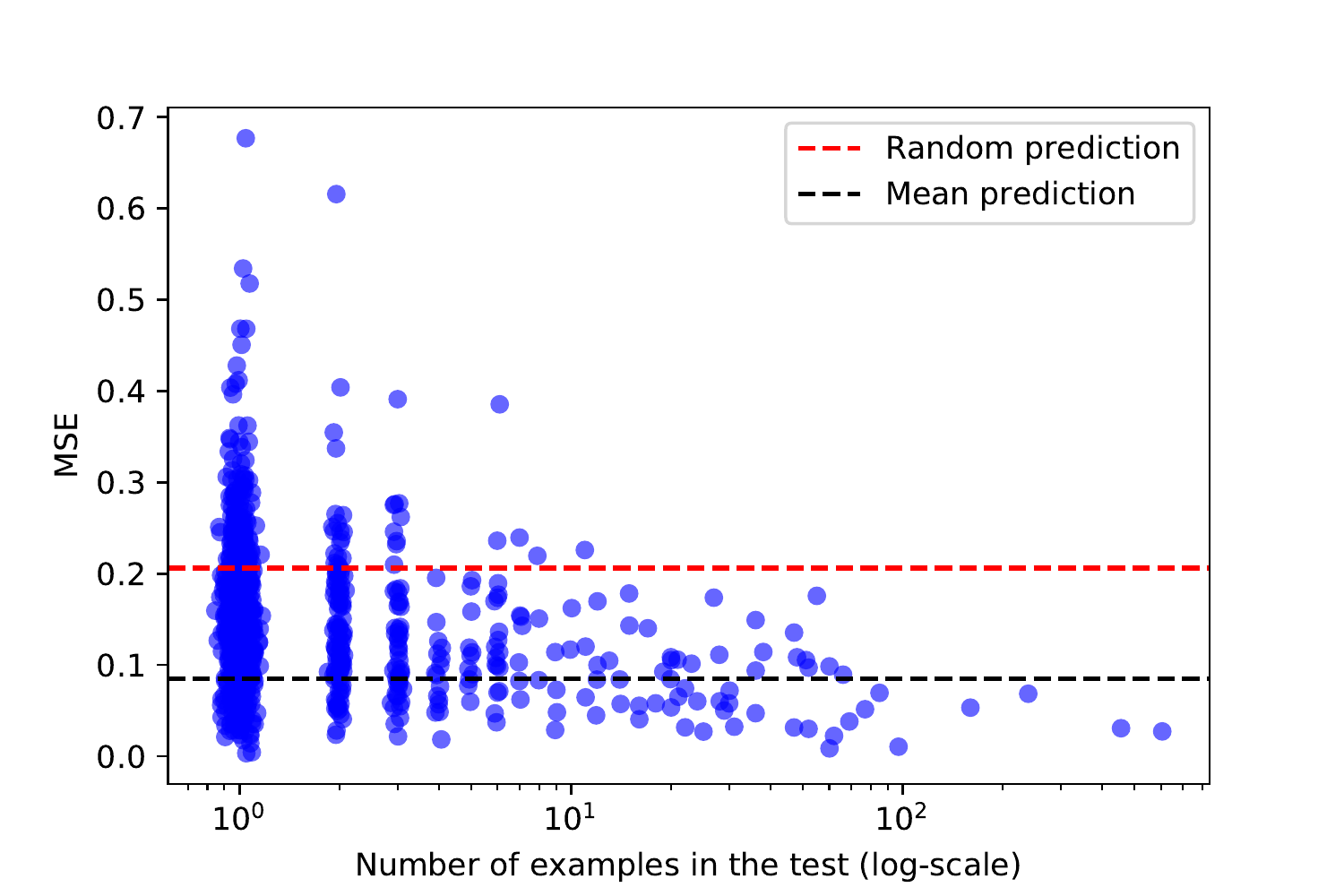}
        \caption{Performance by ligand, split by the number of examples in the set}
    \end{subfigure}%
    
    \caption{Performance of the model detailed in the output space}
    \label{clusters}
\end{figure}

Next, we investigate the behaviour of our model beyond a simple average MSE score. To do so we broke down our results per ligand and clustered the ligands into a dendrogram to measure the effect of the distribution of ligands in the dataset on performance (Figure~\ref{clusters}).

Overall the model performs well across ligand families. The points on the extreme left and right of the dendrogram are ligands that lie far from the main clusters are thus sparsely populated. {\bf Fig. ~\ref{clusters}} illustrates, naturally, that performance improves as more binding sites for a ligand are obtained. 
Regardless, we still have a better prediction than random for most points, including those with few or even one example. 

\subsection{Ligand space reduction to enhance structure-based sampling}

\begin{figure}[H]
    \centering
    \begin{subfigure}[t]{0.45\textwidth}
        \centering
        \includegraphics[scale=0.4]{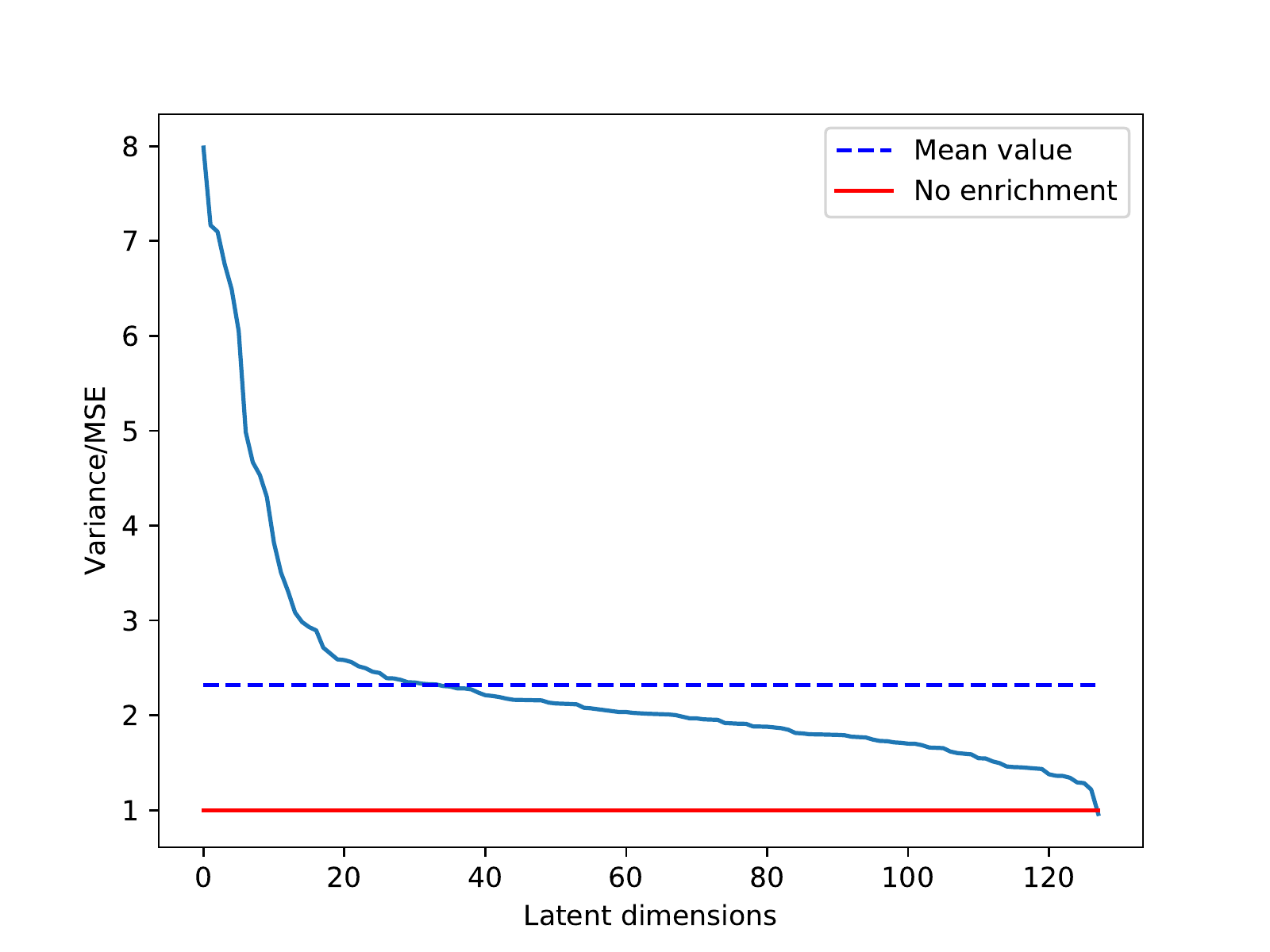}
        \caption{Variance reduction in the latent space per latent dimension}
        \label{varmse}
    \end{subfigure}
    ~ 
    \begin{subfigure}[t]{0.45\textwidth}
        \centering
        \includegraphics[width=\textwidth]{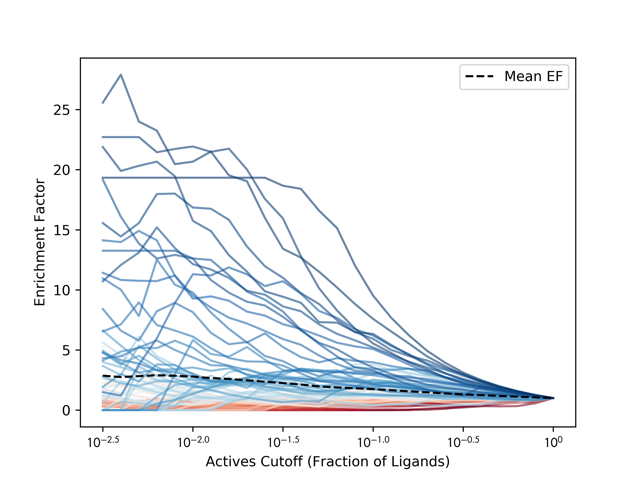}
        \caption{Enrichment factor at different thresholds for each target of the DUDE Database}
    \end{subfigure}%
    \label{VAR_DUDE}
\end{figure}

We now use the average MSE for each output dimension to determine which dimensions are better predicted and as a rough estimate of the uncertainty bounds of our method.
One can compare the MSE per-dimension to the variance of the data (the MSE always returning the average). 
If the MSE is small compared to the variance, we can say that the model has learned to predict this dimension reasonably well (Figure~\ref{varmse}). 

For most dimensions, the reduction is near two-fold over random, yet roughly 30 dimensions experience a reduction of up to 8-fold. 
To contextualize the usefulness of this reduction, we can think of structure-based approaches as taking samples in the chemical space and docking them against the binding site by assessing their affinity. 
The variance of this sampling is reflected by the variety of compounds observed in the PDB. 
The MSE defines a high dimensional box in the latent space, where the affinity is high. 
If we consider uniform sampling the latent space, we would need an expected $10^{42}$ samples (computed value by the product of bins possibility in each dimension) to have a point fall in this box. 
while the usual number of compounds that can be handled by structure-based methods is usually only $10^8$ molecules. 
Due to the curse of dimensionality, even though the result is only giving on average a 2-fold enhancement in each dimension, our method is theoretically able to suggest a point unreachable with the usual structure-based approaches.

\subsection{Case study and enrichment factor}

In this part, we use DUDE \cite{Mysinger2012} as an external database and evaluate the performance of our techniques w.r.t. the enrichment factor. Although we do not aim to match the performance of well-established methods with this settings, it offers an interesting perspective of the potential of our approach.
Ligand-based methods look for similarity between actives compounds that exist in the data set, while structure-based methods explicitly evaluate all candidates. By contrast, we do only one prediction. A more complete validation of the quality of our predictions would be to probe their neighborhood. Here, we are conducting two experiments. First, we want to know if we can correctly identify the sub-region of space with given properties. 
Next, we aim to determine if this sub-region is closer to active compounds than decoys. 
We make our prediction using the 3D structure of the binding site and compute the distance to our prediction. 
To get an EF score, we then sort actives and decoys compounds by distance to this prediction.

Our first results shows that the average distance from a prediction to its associated ligand has an 'MSE' value of 0.13, which is lower than the random one of 0.2. We conclude that our methods are able to identify regions of interest for this task. Importantly, we get an enrichment for most targets, which means that our model is able to make a prediction closer to the region of actives compounds. However, it seems that inactive compounds do not lie in a well separated space. Thus, the closest points of the predictions necessarily include decoys. A strategy to address this caveat would be developing other embeddings that separates compounds based on affinity.

\section{Conclusion and future work}
In this paper, we show that learning a representation of known binding sites enables us to predict sub-regions of the ligand space with improved binding potential.  
These results support more extensive use of binding site information within ligand generative models and suggest novel avenues for improvement for the computational drug discovery pipelines.
Additional binding information such as experimental affinity could further enhance results and help overcome bias of learning only on co-crystallized complexes.

The alignment step within binding sites that helps us to amplify the signal, is a simple yet promising solution to represent 3D structures. Eventually, the binding site representations could be improved through a self-supervised pipeline leveraging the increasing amount of structural data available for proteins.

Finally, several lines of work are already developed to improve generative models for ligands. In this context, our method offers promising perspectives to generate   relevant seed and help for further optimization~\cite{griffiths2017constrained}.

\subsection*{Code and Setup} \label{implementation}
We used a standard model with 7 convolution layers and 2 fully connected. The detailed architecture can be found on \href{https://github.com/Vincentx15}{GitHub} and is described in the \texttt{models/SmallC3D.py} class. One can also run \texttt{python main.py --summary} to get a summary of the model architecture and parameters that were used. Training was performed on 4 NVIDIA P100 Pascal GPU using 20 Broadwell cores.

\subsection*{Acknowledgements}

We thank Mathieu Blanchette and William Hamilton for comments on the text and useful discussion. We thank Robin Winter for his help using and implementing the cddd project. We thank Mario Geiger and Tess Schmidt for advice and help implementing the se3cnn pipeline. 

\newpage
\nocite{*}
\bibliographystyle{unsrt}
\bibliography{citations}

\end{document}